\documentclass[aip,jap,preprint,graphicx,amsmath,amssymb]{revtex4-1}
\usepackage{graphicx}
\usepackage{epstopdf}
\usepackage{amsmath}
\usepackage{amsfonts} 
\usepackage{dcolumn}% Align table columns on decimal point
\usepackage{bm}% bold math

\draft % marks overfull lines with a black rule on the right

\begin{document}
%\preprint{AIP/123-QED}
% Use the \preprint command to place your local institutional report number 
% on the title page in preprint mode.
% Multiple \preprint commands are allowed.
%\preprint{}

\title{Neutral-cluster implantation in polymers by computer experiments} %Title of paper

% repeat the \author .. \affiliation  etc. as needed
% \email, \thanks, \homepage, \altaffiliation all apply to the current author.
% Explanatory text should go in the []'s, 
% actual e-mail address or url should go in the {}'s for \email and \homepage.
% Please use the appropriate macro for the type of information

% \affiliation command applies to all authors since the last \affiliation command. 
% The \affiliation command should follow the other information.

\author{Roberto Cardia}
%\email[]{Your e-mail address}
%\homepage[]{Your web page}
%\thanks{}
%\altaffiliation{}
\affiliation{Dipartimento di Fisica Universit\`a di Cagliari, Cittadella Universitaria, I-09042 Monserrato (Ca), Italy}
\author{Claudio Melis}
\affiliation{Dipartimento di Fisica Universit\`a di Cagliari, Cittadella Universitaria, I-09042 Monserrato (Ca), Italy}
\author{Luciano Colombo}
\affiliation{Dipartimento di Fisica Universit\`a di Cagliari, Cittadella Universitaria, I-09042 Monserrato (Ca), Italy}
% Collaboration name, if desired (requires use of superscriptaddress option in \documentclass). 
% \noaffiliation is required (may also be used with the \author command).
%\collaboration{}
%\noaffiliation

\date{\today}

\begin{abstract}
In  this work we  perform atomistic model potential molecular dynamics  simulations by means of state-of-the art force-fields to study the implantation of a single Au nanocluster on a Polydimethylsiloxane substrate. 
All the simulations have  ben performed on realistic substrate models containing up to $\sim$ 4.6 millions of atoms having depths up to $\sim$90 nm and lateral dimensions up to $\sim$25 nm. We consider both entangled-melt and cross-lin ed  Polydimethylsiloxane amorphous structures.

We show that even a single cluster impact on the Polydimethylsiloxane substrate remarkably changes the polymer local temperature and pressure. Moreover we observe the presence  of craters created on the polymer surface having lateral dimensions comparable to the cluster 
radius and depths strongly dependent on the implantation energy. 

Present simulations suggest that the substrate morphology is largely affected by the cluster impact and that most-likely such modifications favor the the penetration of the next  impinging clusters.
\end{abstract}

% insert suggested PACS numbers in braces on next line

\maketitle %\maketitle must follow title, authors, abstract and \pacs

% Body of paper goes here. Use proper sectioning commands. 
% References should be done using the \cite, \ref, and \label commands
\section{Introduction}
Electronic devices fabricated on bendable plastic substrates are widely believed to have great potential for applications in electronics. Much progress has been made in this area, in particular in the fabrication of  devices and circuits, ranging from electronic paper-like display devices \cite{Lee2006, Rogers2001,Gelinck2004} to sensor skins \cite{Someya2005}, circuits suitable for radio frequency identification tags \cite{Baude2003,Cantatore2007}, non-volatile memories \cite{Ouyang2004}, photovoltaics \cite{Ouyang2004,Mayer2007} , smart clothing \cite{Schubert2006}, and  actuators \cite{Engel2006, Rosset2009}. 

Stretchable electronics represents a much more challenging class of systems \cite{Kim2008} with applications mainly in the biomedical area, where circuits must be integrated with biological tissues. Stretchability demands that the circuits have the capacity to absorb large strain without degradation in their electronic properties.
Materials used in stretchable electronics range from entropy elastic elastomers to energy elastic solids. Stretchable circuits are made of diverse
materials that span more than 12 orders of magnitude in Young elastic modulus.

In this work we focus, in particular, on stretchable electrodes, formed by metallic paths inserted into elastomeric substrate, mainly fabricated via the direct metallization of elastomers\cite{Whitesides2006}.
Polydimethylsiloxane (PDMS)  is  the most used elastomer substrate because  it combines biocompatibility with suitable mechanical properties\cite{Whitesides2006}. Moreover, PDMS can be easily synthesized  by available laboratory techniques and it is not attacked by most process chemicals\cite{Mandlik2008}.
Amorphous PDMS can be synthesized in different microscopic structures, depending on the applications, including entangled-melt (EM) films  (composed by a network of chains strongly wrapped around each other) as well as  cross-linked  (CL) structures  in which individual chains are chemically bonded to each other through suitable cross-linker molecules\cite{Dollase2002} .

The metallization of PDMS to produce micrometric conductive pathways is typically obtained by  metal vapor deposition\cite{Graz2009} or by metal ion implantation\cite{Maggioni2004}. 
One of the main drawbacks of  the metal vapor deposition technique  is the delamination of the conducting layers even at very low deformations.
This problem is solved by metal ion implantation, which provides improved adhesion between the conducting layers and the polymeric substrate. In this technique nobel metal ions are implanted with energies in the range of  the KeV/atom\cite{Maggioni2004,Rosset2009}, thus forming a conductive layer just below the substrate surface. However,  it has been shown that upon the ion implantation the insulating polymeric substrate is likely  degraded  by carbonization and the presence of charges\cite{Ravagnan2009}.

Recently a novel implantation technique named  "Supersonic cluster beam implantation" (SCBI) has been proposed \cite{Ravagnan2009,Corbelli2011}. SCBI consists in pointing a collimated beam of neutral metallic clusters  towards a polymeric substrate. The size distribution of the clusters ranges from $3$ nm to $10$ nm and the typical implantation kinetic energy is about $1.0$ eV/atom.
Even if the kinetic energy comparatively lower than in the ion implantation technique, the neutral clusters are able  to penetrate up to tens of nm into the polymeric target forming a buried conducting  layer and  avoiding charging and carbonization.
The mechanism driving SCBI is the so-called "clearing-the-way" effect \cite{CLW,CLW1,CLW2}, namely the front atoms in the cluster collide with target atoms conveying them sufficient momentum to clear the way for the following cluster atoms. As a consequence, the low-energy cluster penetration depth is increased with respect to the high-energy single ion penetration. The "clearing the way" effect has been confirmed and formalized by classical molecular dynamics in which  Ag$_n$ clusters (n = 20-200) impacted with total energies in the range of 0.5-6 keV into graphite\cite{CLW1}. The penetration depth $P_d$ was related to the implantation energy $E_0$, the cluster number of atoms $n$, and the substrate  cohesive energy $U$  according to the following scaling law\cite{CLW}:
\begin{equation}\label{CTW}
P_d\propto\frac{E_0n^{1/3}}{U}
\end{equation}
The underlying assumption is that that $E_0$ is mainly spent in breaking bonds of the substrate while penetrating into it. Therefore, the penetration depth turns out to be inversely proportional to the cohesive energy U of the substrate.

The SCBI technique has been recently applied using Au-nanoclusters (Au-nc) implanted on a PDMS substrate\cite{Corbelli2011}. The resulting nanocomposite can withstand many deformation cycles, decreasing their electrical resistance upon cyclical stretching\cite{Corbelli2011}.
While the effectiveness of this novel technique has been demonstrated\cite{Corbelli2011}, the physical and chemical phenomena underlying the SCBI process at the microscopic scale remains to be fully understood. As a matter of fact, the implantation process of neutral metal clusters in a polymer matrix is a problem rarely studied in literature \cite{Ravagnan2009, Corbelli2011,Marelli2011} and the mechanical and electrical properties of the resulting nanocomposites need to be fully characterized at the microscopic scale. 
In particular, it is important to investigate the polymer microstructure evolution in terms of temperature and pressure waves generated  upon implantation and the resulting polymer surface damage. 
Another important property to characterize  is the dependence of the cluster penetration depth on the implantation energy and cluster dimension. This information is relevant to better control the SCBI process since it allows to tune the penetration depth by changing the implantation energy and/or the cluster dimensions. Finally, it is important to assess the different polymer response upon the implantation depending on its microscopic structure, namely depending on the entangled-melt vs. cross-linked bond network.

In this perspective, computer simulations represent a key tool in order to fully describe at the atomistic scale both the SCBI process and the following microstructure evolution of the substrate.
In this work we present molecular dynamics (MD) simulations of the implantation of a single Au cluster on a PDMS substrate. We understand that this single event (although it can hardly be compared to the full SCBI experiment) is the most fundamental process ruling SCBI and, therefore, we believe it deserves a careful investigation. To the best of out knowledge a direct simulation of such a cluster impact is still missing in Literature.

In order to keep the present simulations on the actual experimental length-scale we make use of very large-scale MD simulations by aging as many as $\sim$ 4.6 millions atomic trajectories. Furthermore, special care is played to carefully model the cluster-substrate interactions, as well as the possible presence of molecular linkers into the PDMS film. To this aim, we make use of a state of the art force-field which has been accurately benchmarked on well-known physical properties of pristine PDMS.

The goal of the present work is three-fold:(i) modeling at the atomic scale the impact (and following penetration) of the cluster into the substrate; (ii) linking the penetration depth to the implantation parameters (like e.g. the cluster implantation energy or size, and the structure of the substrate); (iii) characterize the surface damage upon impact.\\
The much more challenging full simulation of the SCBI process (i.e. the simulation of a multi-impact cluster deposition) will be investigated in a following work.
%\label{}
\section{Theoretical framework} 
 \subsection{The force field}
\begin{figure}[h]
\includegraphics{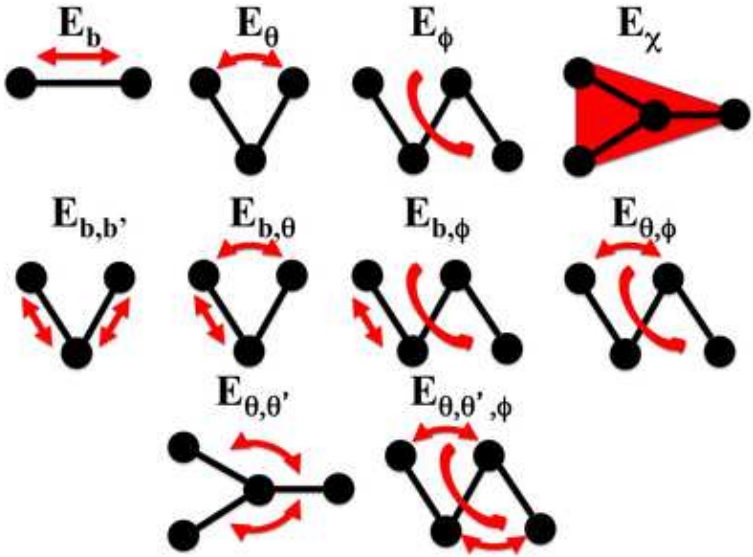}
\caption{Graphic illustration of the COMPASS force-filed terms.
  \label{compass-fig}}.
\end{figure}

 All simulations are performed using the COMPASS (condensed-phase optimized molecular potentials for atomistic simulation studies) force-field \cite{Sun1998},  including off-diagonal cross-coupling terms and high-order (cubic and quartic) force constants. The parametrization of the COMPASS  force-field allows to describe, using the same functional form, organic (single molecules and  polymers) as well as  inorganic materials \cite{Sun1998, Heinz2008}. 
 The functional forms used in this force field are the same as those used in consistent force-fields (CFF)\cite{CFF}:	
 \begin{equation}\label{COMPASS}
 E_{\mathrm{total}} = E_b + E_{\theta} + E_{\phi} + E_{\chi} + E_{b,b'} + E_{b,\theta} + E_{b,\phi} +  E_{\theta,\phi} + 
E_{\theta,\theta'} +  E_{\theta,\theta',\phi} + E_{q} + E_{\mathrm{vdW}}
\end{equation}
where:
\begin{equation}\label{COMPASS1}\begin{split}
E_b = &\sum_b \left[ k_2 (b-b_0)^2 + k_3 (b-b_0)^3 + k_4 (b-b_0)^4   \right]\\
E_{\theta} = &\sum_{\theta} \left[ k_2 (\theta-\theta_0)^2 + k_3 (\theta-\theta_0)^3 + k_4 (\theta-\theta_0)^4   \right]\\ 
E_{\phi} =& \sum_{\phi} \left[ k_1(1-\cos \phi) + k_2(1-\cos 2\phi) + k_3(1-\cos 3\phi) \right]\\
 E_{\chi} =& \sum_{\chi} k_2 \chi^2, E_{b,b'} = \sum_{b,b'} k(b-b_0)(b'-b_0')\\
E_{b,\theta} =& \sum_{b,\theta} k(b-b_0)(\theta-\theta_0)\\
\end{split}\end{equation}
\begin{equation}\label{COMPASS3}\begin{split}
E_{b,\phi} = &\sum_{b,\phi} (b-b_0) \left[k_1 cos \phi + k_2 cos 2\phi + k_3 cos 3\phi\right]\\
E_{\theta,\phi} = &\sum_{\theta,\phi} (\theta-\theta_0) \left[k_1 cos \phi + k_2 cos 2\phi + k_3 cos 3\phi\right]\\
E_{\theta,\theta'} =& \sum_{\theta,\theta'} k(\theta-\theta_0)(\theta'-\theta_0')\\
E_{\theta,\theta',\phi} =& \sum_{\theta,\theta',\phi} k(\theta-\theta_0)(\theta'-\theta_0')\cos \phi\\
E_{q}=&\sum_{ij} \frac{q_i q_j}{r_{ij}}
\nonumber \end{split}\end{equation}
and
\begin{equation}\label{COMPASS2}
E_{\mathrm{vdW}}= \sum_{ij} \epsilon_{ij} \left[ 2 \left(\frac{r_{ij}^0}{r_{ij}} \right)^9 -3  \left(\frac{r_{ij}^0}{r_{ij}} \right)^6\right]
\end{equation}
with the following combining rules
\begin{equation}\label{COMPASS3}
r_{ij}^0 = \left( \frac{ (r_i^0)^6 + (r_j^0)^6 }{2} \right)^{1/6}
\epsilon_{ij} = 2 \sqrt{\epsilon_i \cdot \epsilon_j} \left( \frac{ (r_i^0)^3 \cdot  (r_j^0)^3 }{ (r_i^0)^6 \cdot  (r_j^0)^6 }  \right)
\end{equation}
The functions can be divided into two types: valence terms including diagonal ($E_{b,b'}$, $E_{\theta,\theta'}$) and off-diagonal cross-coupling terms ($E_{b,\theta}$, $E_{b,\phi}$, $E_{\theta,\phi}$, $E_{\theta,\theta',\phi}$), and nonbond interaction terms. The valence terms represent internal coordinates of bond ($b$), angle ($\theta$), torsion angle ($\phi$), and out-of-plane angle ($\chi$), while the cross-coupling terms include combinations of two or three internal coordinates (see \ref{compass-fig}). The cross-coupling terms are important for predicting vibration frequencies and structural variations associated with conformational changes. In \ref{COMPASS1} the subscript $0$ denotes the reference values of the bond, angle, dihedral angle and out-of-plane angle, while $k$, $k_1$, $k_2$, $k_3$, $k_4$ are the force constants determined from the quantum mechanical energy surface. Terms involving explicit internuclear distances, $r$, represent nonbond interactions which are composed by  a $E_{vdW}$ Lennard Jones (LJ) 9-6 function for the van der Waals (vdW) term and a $E_{q}$ Coulombic function for an electrostatic interaction. The LJ-9-6 function is particularly suitable for the case of cluster impacts with respect to LJ-12-6 function, which is known to be too hard in the repulsion region.
 The COMPASS force field parameters  have been optimized for PDMS by fitting to $ab$ $initio$ calculations\cite{Sun1997, Sun1995}.
 \subsection{Entangled-melt PDMS}
\begin{figure}[h]
\includegraphics{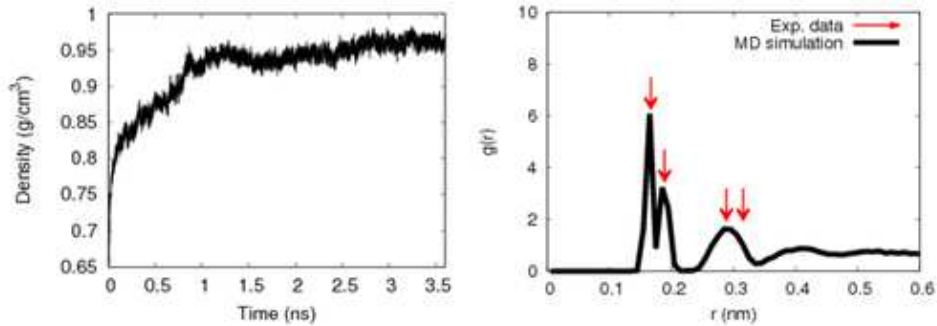}
\caption{Left: EM-PDMS density evolution during the 3.5ns-long NPT equilibration. Right: Calculated final radial distribution function. Arrows mark the coordination peaks obtained by  experimental X-ray diffraction\cite{PDMS-AMORPH3,PDMS-AMORPH4}.
  \label{exp}}.
\end{figure} 
We generate amorphous (a) EM-PDMS  by means of the  Theodorou e Suter \cite{AMORPH,AMORPH1} algorithm  implemented in the  Materials Studio \cite{MATERIALSTUDIO} package. The simulation cell is composed by $6$ chains  with $100$ PDMS monomers each, having in total $6462$  atoms  and an initial simulation cell volume is set as to correspond to a density of 1 g/cm$^3$. 
After the  EM-PDMS structure is generated, we equilibrate the cell volume using the LAMMPS package by performing a constant-temperature constant pressure (NPT) simulation for 3.5 ns at a pressure of 1 Atm  and a temperature of 300K. The final density of 0.96 g/cm$^3$ (reported in \ref{exp} (left)) is in good agreement with the experimental data for PDMS at 300 K \cite{PDMS-AMORPH3,PDMS-AMORPH4,PDMS-AMORPH1,PDMS-AMORPH2}. 
We calculate the corresponding radial distribution function $g(r)$  (see \ref{exp}, right) finding a quite good agreement with available experimental data\cite{PDMS-AMORPH3,PDMS-AMORPH4}.
% both in terms of peak positions. Differences are found on the first peak intensity and on the peak positions above 0.3 nm.
%Once the AEM-PDMS unit cell was equilibrated, we obtained the whole PDMS substrate by replicating the equilibrated unit cell.
\cite{PDMS-AMORPH3,PDMS-AMORPH4}.
\subsection{Crossed-linked PDMS}
\begin{figure}[h]
\includegraphics{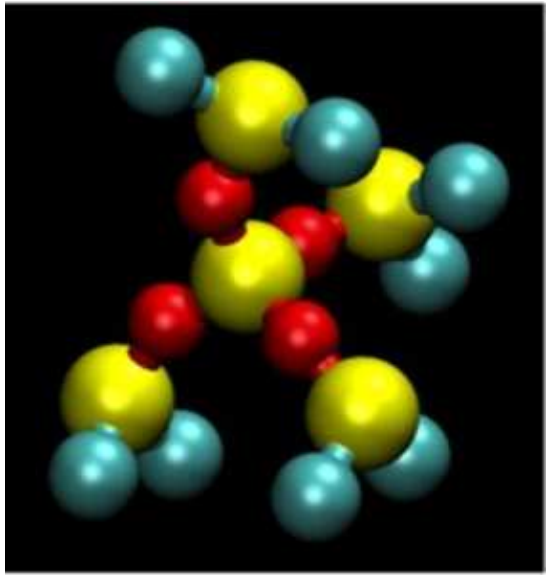}
\caption{Balls and stick representation of the tetrakis(dimethylsiloxy)silane molecule. Carbon atoms are shown in cyan, Oxygen atoms in red and Silicon atoms in yellow.The hydrogen atoms have been removed for clarity.
  \label{tetrakis}}.
\end{figure}

The cumputer generation of cross-linked polymers  is computationally challenging due to the need of simulating  the chemical reactions involving the creation of covalent bonds between the cross-linker molecule and PDMS.
For this reason only few attempts have been made in order to describe the PDMS cross-links reactions via MD simulations\cite{Chenoweth2005, Shemella2011,Lacevic2008, Shernella2011,Heine2004}
% Chenoweth2005, Lacevic2008 }.
%Among these different simulation protocols we follow the work of Heine at al.\cite{Heine2004} in which
 In this work the tetrakis(dimethylsiloxy) silane was used as cross linker (\ref{tetrakis}) similarly to Ref. \citep{Heine2004}. Tetrakis(dimethylsiloxy) silane is composed by  1 central and 4 peripheral  Silicon atoms, 4 Oxygen atoms, 8 Carbon atoms and 24 Hydrogen atoms.
In the presence of a catalyst, each tetrakis(dimethylsiloxy)silane cross-linker is able to bond to four terminal hydroxy groups on the PDMS chains via its 4 peripheral  Si atoms. We simulate this mechanism by allowing the covalent bonds between pairs of atoms to dynamically create, during the MD simulation, according to specified criteria. In  particular, the cross-linking reaction is mimicked by forming a bond between one of the 4 terminal peripheral Si on the cross-linker and the nearest terminal O on the PDMS chain. The bond forms with unit probability when the Si and O  atoms are separated  by a distance  less than  0.1 nm.
According to the experimental data\cite{MILANI}, we use a 8 $\%$ cross-linkers weight occurrence, corresponding to 7200 cross-linker  molecules. In order to reduce the computational cost of the PDMS cross linking procedure, we initially work on the previously equilibrated EM-PDMS  building-block (see previous section)  by adding 10 cross-linker molecules (corresponding to a 8 $\%$ occurrence). After this preparatory step, we replicate the so obtained building-block (see below) in order to generate the whole PDMS substrate.
At this point we performed a NVT simulation at T=500 K for 0.1 ns in which the cross-linker molecules and PDMS were allowed to dynamically bond according to the criteria specified above. After 0.1 ns more than  90$\%$ of cross-linkers were bonded to PDMS.

All the simulations are performed by using the LAMMPS code \cite{PLIMPTON1995}. The velocity-Verlet algorithm with a time step of 1.0 fs  is used to solve the equations of motion. A  particle-mesh Ewald algorithm is used for the long-range electrostatic forces, and the Van der Waals interactions were cut off at 0.1 nm. The Nos\'e  -Hoover thermostat/barostat with corresponding relaxation times equal to 100 and 500 fs are used to control the simulation temperature and pressure, respectively.
 \subsection{Metal cluster}
As for the description of the Au-nc, we also use the COMPASS force-field, and in particular the $9-6$ Lennard-Jones potential. In fact face-centered cubic (fcc) elemental metals can be successfully described by means of suitable Lennard-Jones potentials\cite{Heinz2008}. The COMPASS $9-6$ Lennard-Jones $\epsilon$  and $r_0$ parameters for Au have been recently further optimized  in order to describe:  densities, surface tensions, interface properties with organic molecules. All these properties are found  in good agreement with experiments performed  under ambient conditions\cite{Heinz2008}.

 \subsection{Simulation procedure}
 Once the PDMS building-block (both EM and CL) is equilibrated we obtain the  PDMS substrate for the SCBI simulation by the following procedure:
\begin{enumerate}
\item We replicate  $N$ x $L$ x $M$ times (where N, L, and M are integers)  the basic building-block along the $x$, $y$ and $z$ directions respectively. %Once the cell architecture have been selected, we performed the equilibration procedure before the cluster impact.
We then equilibrate such a PDMS bulk sample for $0.2$ ns  at a temperature  T=500 K under orthorhombic periodic boundary conditions along the $x$, $y$ and $z$ directions (see \ref{CELL-ARCH}). This annealing step is crucially important in order to remove possible artifacts deriving from  the previous fictitious periodicity imposed by the replica of the PDMS basic building-block. 
\item Once the bulk PDMS sample is equilibrated, we create the PDMS surface by removing the periodic boundary conditions along the $z$ direction and by  further equilibrating the slab sample for 0.1 ns at T=300K . While the topmost surface is let to freely age under any following MD run, the bottom surface is kept fixed in order to emulate a macroscopically thick experimental sample.
\end{enumerate}
\subsection{Simulation cell architecture}
\begin{figure}[h]
\includegraphics{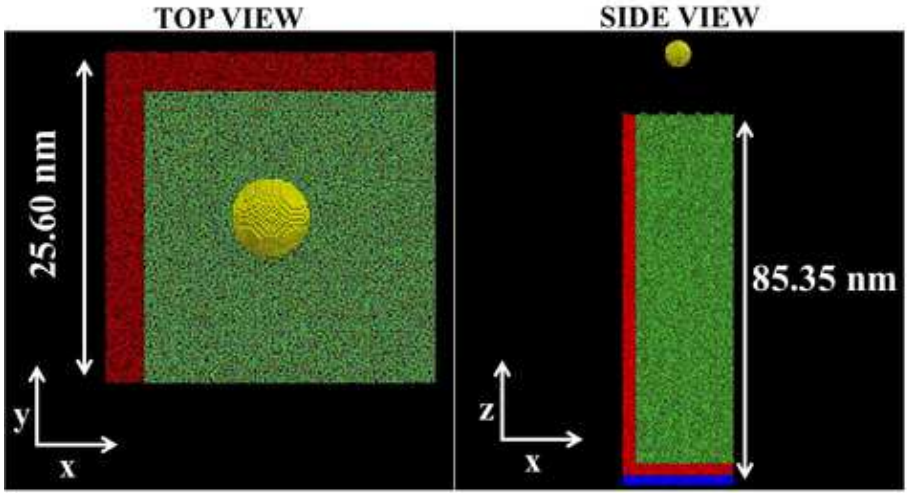}
\caption{Top (left) and side (right) view of the simulation cell. The fixed  bottom region is shown in blue, the frame region coupled to a thermostat  is shown in red, while  the rest of the cell evolves by means of newtonian dynamics.The cluster is shown in yellow.
\label{CELL-ARCH}}
\end{figure}

  One of the main issues related to implantation simulations is due to the presence of the (x,y) periodic boundary conditions that can eventually give
rise to unphysical interactions (i.e. overlap between pressure and temperature waves generated upon the cluster impact) between neighbor cell replica. In order to avoid such artifacts, we create a frame region at the cell boundaries which are coupled to a Nos\'e  -Hoover thermostat.  The role of the frame region is  to mimic a bulk absorbing the excess of heat generated upon the cluster impact.The rest of the simulation cell, where the cluster impact actually takes place, is simulated without any thermostat or barostat. In other words, atomic trajectories in this region are aged according to pure Newton dynamics. The simulation cell architecture is shown in  \ref{CELL-ARCH} where  the fixed bottom region is shown in blue, the frame region coupled to a thermostat (lateral and bottom region) in red, while the rest of the cell evolves by means of newtonian dynamics.

We then  place  the Au-nc with the center of mass at a distance of  $10$ nm from the EM-PDMS surface. According to the SCBI experimental data\cite{MILANI} the cluster vibrational temperature before the impact is $\sim$100 K while the PDMS temperature is  $\sim$300 K. Therefore we equilibrate the Au-nc  at T=100K and the PDMS at 300 K for 0.1 ns before the impact.
% The last part of the simulation was devoted to equilibrate the whole system with the  the cell geometry described in \ref{CELL-ARCH}  in which the PDMS framework  and the Au-nc evolve coupled to a thermostat while the rest of the cell evolves by means of newtonian dynamics.

\subsection{Choice of the substrate dimensions.}
We select the optimal substrate dimensions  by performing several implantation simulations on different EM-PDMS substrates of increasing thickness and section.
\begin{figure}[h]
\includegraphics{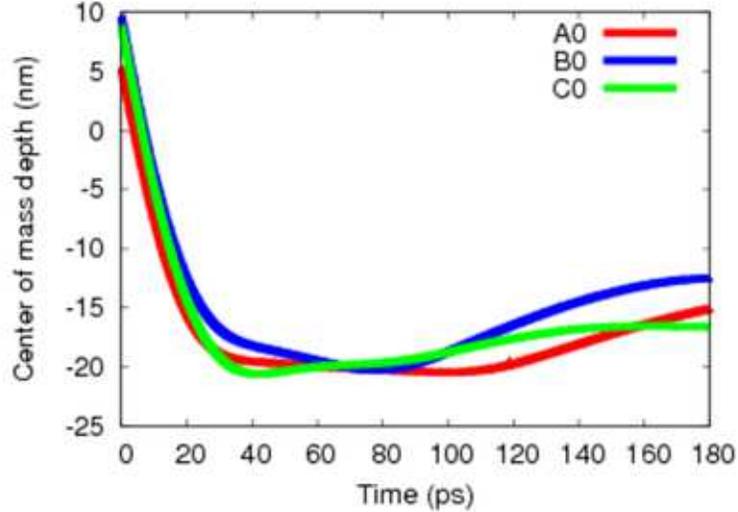}
\caption{Time evolution of the Au-nc penetration depth for 3 EM-PDMS substrates of increasing dimensions corresponding to samples $A_0$, $B_0$ and $C_0$. Contrarily to $A_0$ and $B_0$, the $C_0$ substrate has an optimal lateral dimension and depth that allows  to stabilize the cluster position inside the EM-PDMS after 180 ps\label{PD0}}
\end{figure} 
The implantation simulations are performed by considering Au-nc having a radius of $3$ nm implanted at an energy of $2$ eV/atom along the $z$ direction (see \ref{CELL-ARCH}) .  \ref{PD0} shows the time evolution of  the Au-nc penetration depth during the implantation. The cluster penetration depth at time $t$ is calculated as the difference between the maximum EM-PDMS surface height at $t$=$0$ ps and the cluster center of mass height at time $t$.
We test three different substrate dimensions having different lateral dimensions and depths as shown in \ref{sub1}.
We observed that while the absolute penetration depth is not largely affected by the substrate dimension, the long-time microstructure evolution sizably depends on them. 
%sample $A_0$ ($1923840$ atoms ), having lateral dimensions of  $17.07$ nm $x$ 1$7.07$ nm  and a depth  of $85.35$ nm; sample $B_0$ ($2885760$ atoms ), having the same lateral dimensions as $A_0$ and a depth of $128.01$ nm; sample $C_0$ ($4328640$ atoms), having lateral dimensions of  $25.60$ nm $x$ $25.60$ nm and a depth of $85.35$ nm.
\begin{table}
\begin{ruledtabular}
\begin{tabular}{|c|c|c|c|}
\hline
&Sample A$_0$&Sample B$_0$&Sample C$_0$\\
\hline
x,y (nm)& $17.07$ $x$ 1$7.07$& $17.07$  $x$ 1$7.07$&$25.60$  $x$ $25.60$\\
z (nm) &$85.35$&$128.01$& $85.35$\\
\# atoms&$1923840$&$2885760$&$4328640$\\
\hline
\end{tabular}
\caption{Lateral dimension, depth and total number of atoms of the substrates $A_0$, $B_0$ and $C_0$.\label{sub1}}
\end{ruledtabular}
\end{table}
In particular, while in the case of  $A_0$ and  $B_0$ the cluster position on the PDMS polymer is not stabilized after $180$ ps, the $C_0$ substrate has an optimal lateral dimension and depth that allows to efficiently absorb the pressure wave generated upon the cluster impact stabilizing the cluster position inside the EM-PDMS after 180 ps. For this reason we choose the $C_0$ substrate for the implantation simulations.
The time $t=180$  where all the transient post-implantation phenomena are terminated will be hereafter identified as $t_{end}$.
\section{Results: Implantation of 3 nm Au clusters}
\subsection {Entangled melt vs. cross-linked PDMS}
 After the equilibration of the  PDMS+Au-nc system, we perform three different SCBI simulation, respectively  with implantation energy $0.5$, $1.0$ and $2.0$ eV/atom. We chose Au-nc having a radius of $3$ nm, comparable to the experimental SCBI data ($3-10$ nm)\cite{Corbelli2011}.
As far as concern the implantation energy, besides  $0.5$ eV/atom, which is the actual experimental value, we choose also larger implantation energies ($1.0$ and $2.0$ eV/atom)  to verify the effect of this parameter on the cluster penetration and PDMS response.

\begin{figure}[h]
\includegraphics[scale=1.1]{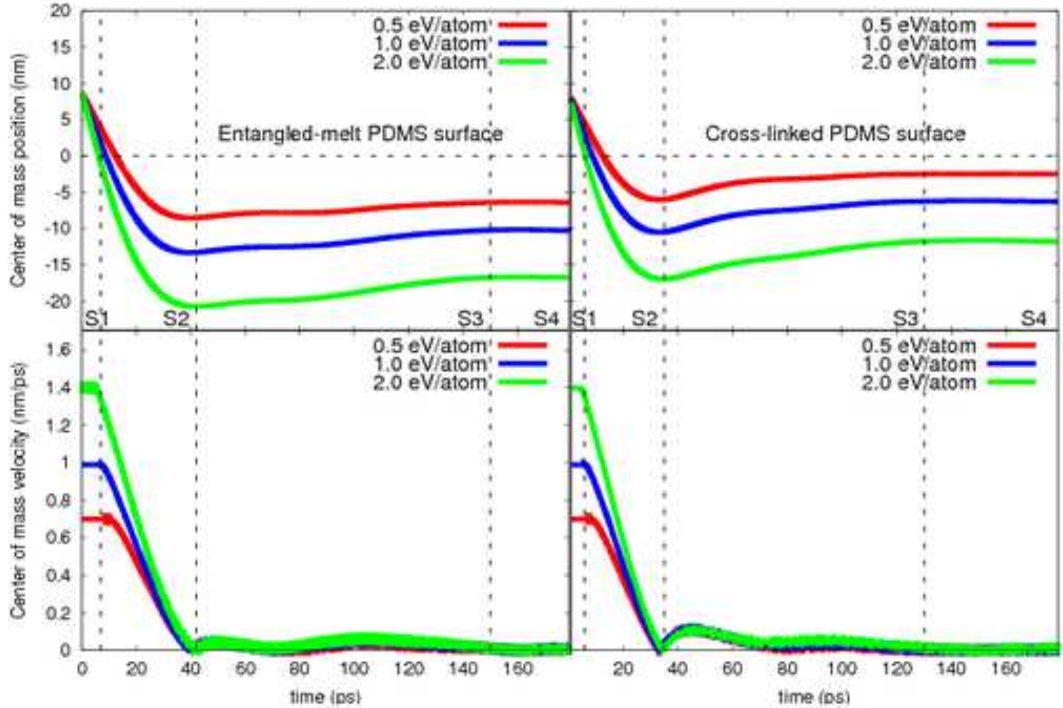}
\caption{Left: position vs. time of the center of mass (top) and corresponding velocity (bottom) 
on the EM-(left) and CL-(right) PDMS for 3 different implantation energies. \label{PD1}}
\end{figure}

 \ref{PD1} shows the time-evolution of the cluster center of mass penetration depth (top)  and the corresponding velocity (bottom) during the simulation, on the EM-(left) and CL-(right) PDMS. 

 Based on \ref{PD1} we can distinguish 4 different microstructure evolution regimes for both EM-and CL-PDMS (hereafter referred to as  $S_1$, $S_2$, $S_3$ and $S_4$  respectively) depending on the polymer response to the cluster implantation. During step $S_1$ the cluster moves in vacuo by linear uniform motion having a constant velocity depending on the implantation kinetic energy.  During step $S_2$  the cluster impacts on the PDMS surface, penetrates inside the substrate and is eventually stopped by friction.
 %The $S_2$ regime last until the cluster is damped due to the substrate friction and the corresponding velocity goes to zero. In the third part of the simulation 
 During step $S_3$ we observe the PDMS response to the cluster impact, at first (cluster penetration) the polymer film undergoes a sizable compression which is then released by its elastic response, able to drag the implanted cluster upward in the polymer. Finally, during step $S_4$ the polymer elastic response is damped and the cluster stabilizes inside the PDMS substrate at a given depth.  These four steps are observed for both EM- and CL-PDMS, independently of the implantation energy, while the respective time duration every regime takes place depends on the initial implantation energy. In particular, we observe that steps  $S_2$ and $S_3$ are longer by increasing the implantation energy, while the cluster stabilization period increases by decreasing the implantation energy.
 
From \ref{PD1} (top panel) we highlight  the differences between the post-implantation evolution in EM- and CL-PDMS. We notice that the cluster penetration depth in EM-PDMS is greater than in the CL-PDMS. This is due to the fact that  in the previous system the polymer chains are bonded only via dispersion and electrostatic interactions, while in the latter one the chains are partially covalently linked via the cross-linker molecule. Since the cluster penetration inside the PDMS matrix involves the breaking of the inter-chains bonds, we conclude that during the penetration inside the EM-PDMS substrate, the cluster experiences a weaker friction with respect to CL-PDMS.
This picture is confirmed by the analysis of the cluster velocity during the implantation (\ref{PD1}, bottom panel). In the case of the CL-PDMS the cluster velocity takes a zero value  earlier ($\sim$ 35 ps) than in the case of EM-PDMS ($\sim$ 40 ps).
\begin{figure}[h!]
\includegraphics{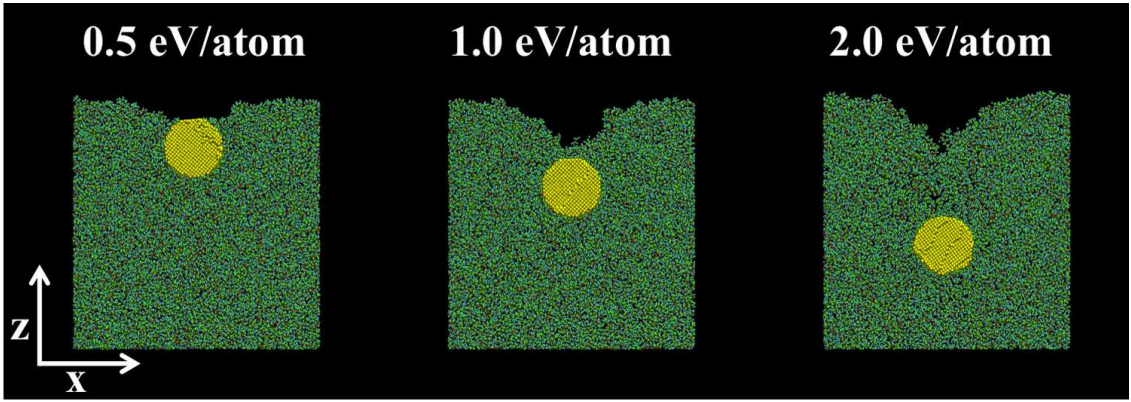}
\caption{Final structures corresponding to $t_{end}$ of the SCBI simulations with implantation energies of $0.5$ eV/atom (left), $1.0$ eV/atom (center) and $2.0$ eV/atom (right). The Au cluster atoms are shown in yellow while the PDMS substrate atoms in green. Only the topmost layer of PDMS is shown.\label{PD2}}
\end{figure} 
We also observe  that the polymer response upon the cluster implantation is more pronounced in the case of CL-PDMS. In fact, we observe a polymer swelling upon the cluster implantation which we quantified as the difference between the polymer maximum height after the implantation at t=$t_{end}$ and before the implantation at $t=0$ ps. In the case of the CL-PDMS we observe a polymer swelling of $\sim$ 3 nm, independently of the implantation energy. On the other hand the EM-PDMS swelling was always less than 1 nm. This difference can be explained by considering that in the case of CL-PDMS  the number of  internal bonds  increases by $\sim$ 8$\%$ due to the presence of the cross linker. Therefore, we expect a comparatively larger elastic response upon the implantation.It is worth remarking that such a swelling phenomena have been indeed observed experimentally\cite{MILANI}.This qualitative agreement between real and simulated SCBI process stands for the overall reliability of the present computational setup.
% which is greater than in the case of EM-PDMS.The PDMS swelling upon Au cluster implantation has been experimentally observed\cite{MILANI}. 

\ref{PD2} shows the final structures at $t_{end}$ of a (x,y) EM-PDMS section having depth of 2 nm and height of 30 nm (similar structures were observed for CL-PDMS). This cross-view confirms that the cluster penetration depth increases with the implantation energy, but also  adds valuable new information, namely:  the polymer surface is strongly damaged upon the cluster implantation (in particular for the implantation energy of 2 eV/atom). We observe the presence of a crater created on the PDMS surface upon the cluster implantation. The lateral dimensions of the crater are almost unchanged by increasing the implantation energy, while its depth strongly depends on the implantation energy as well as the surface roughness. This feature will be more extensively investigated in a following paper\cite{milani1}.

\begin{figure}[t!]
\includegraphics{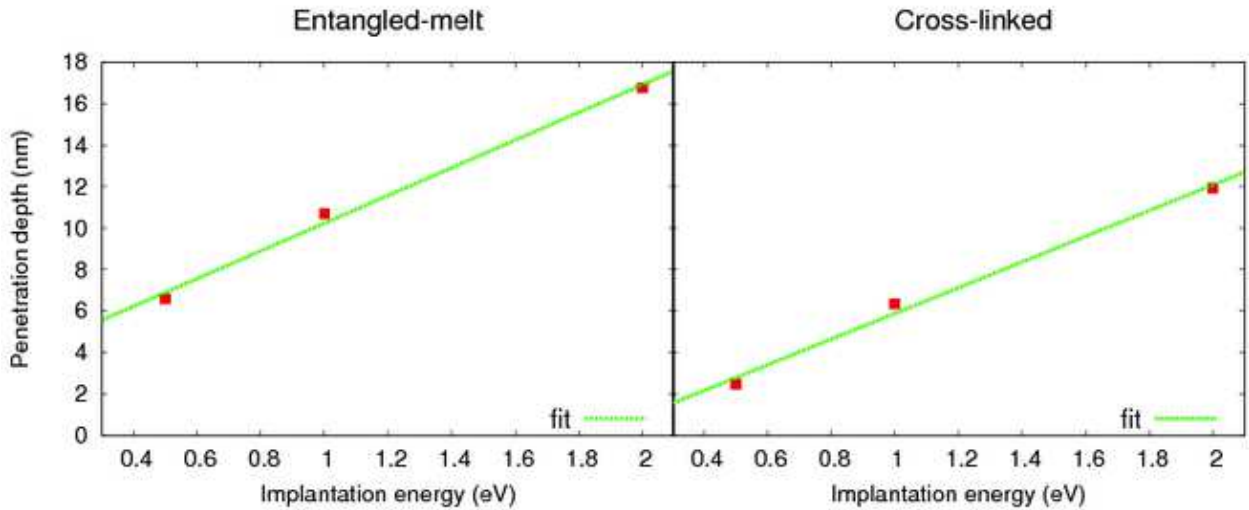}
\caption{3nm cluster penetration depth as a function of the implantation energy (red crosses) fitted with a line (green)  having angular coefficient of  $7$ and $6$ nm/eV for EM-(left) and  CL- (right) PDMS respectively.\label{PD3}}
\end{figure} 
In  \ref{PD3} we plot the cluster penetration depth vs. the implantation energy in the case of EM-(left) and CL-(right) PDMS. As predicted by \ref{CTW}, a linear relationship is found which we easily fit finding that by increasing the implantation energy by $1$ eV/atom cluster penetration depth increases by $7$ nm and $6$ nm for EM-and  CL-PDMS, respectively. This information is useful for the SCBI experiments since it allows to tune the cluster penetration depth by changing the cluster implantation energy. We further remark that the internal cohesive energy contribution $U_{CL}$ in the case of CL-PDMS is 1.2 times larger than EM-PDMS. Therefore we expect a corresponding decrease in in the cluster penetration depth $P_d$ as indeed referred in  \ref{CTW}.

Besides the cluster penetration depth it is important to characterize the temperature and pressure waves generated inside the PDMS substrate upon the cluster implantation. For this reason we calculate the substrate temperature during the implantation as a function of the PDMS depth.
\begin{figure}[t!]
\includegraphics{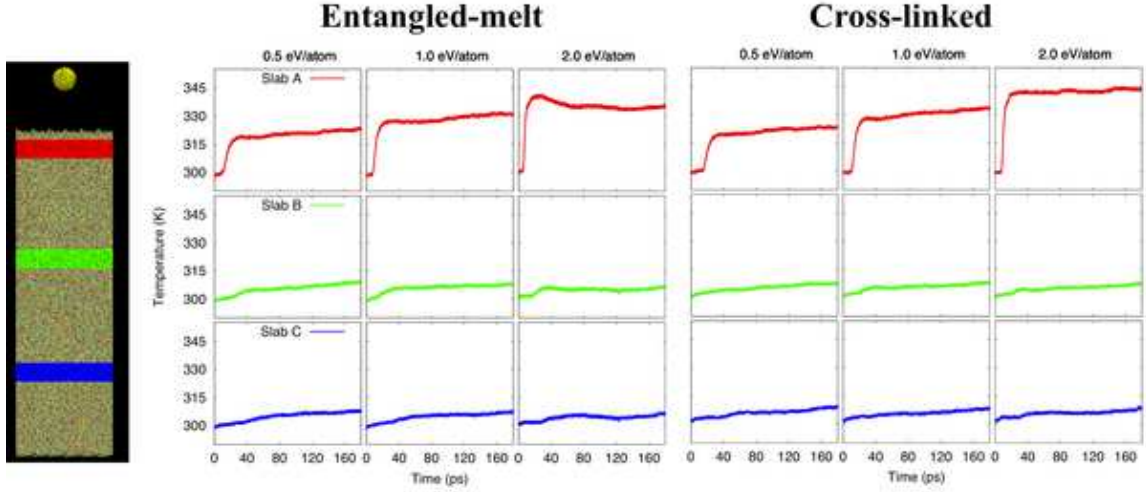}
%\caption{Temperature field (vs. time and  depth) of the EM-(left) and CL-(right) PDMS for the implantation energies of 0.5 eV (top), 1.0 eV (middle) and 2.0 eV (bottom).\label{PD4}}
\caption{Temperature time evolution of the EM- and CL-PDMS for the implantation energies of 0.5 eV/atom, 1.0 eV/atom and 2.0 eV/atom. The first, second and  third  lines  represents the temperature evolutions inside the $A$, $B$ and $C$ slabs respectively.\label{PD15}}
\end{figure} 
\ref{PD15} shows the temperature of  three slabs ($A$,$B$,$C$) for the EM- AND CL-PDMS samples during the 180 ps -long simulation. We choose slab $A$ close to the EM-PDMS surface(70-20 nm depth), slab $B$ corresponding the middle of the substrate (370-320 nm depth), and slab $C$ close to the  PDMS bottom (670-620 nm depth). 
We observe a sudden temperature increase upon the implantation for both EM- and CL-PDMS samples. In particular the overall temperature increase corresponding to the slab $A$ close to the surface ($\Delta T_{A}$)  goes from $\sim$20 K to $\sim$50 K by increasing the implantation energy. This temperature wave propagates inside the substrate being strongly dumped for  deeper slabs  ($\Delta T_{B}$ $\sim$10 K and $\Delta T_{C}$ $\sim$5 K). At $t_{end}$ we  observe two main temperature regions present in the polymer, namely: a "hot" region ($H$) at temperature in the range $\sim$ 320-350 K  close to the surface and another "cold" region at temperature in the range of $\sim$ 300-310 K at larger depth. We observe that the overall temperature increase is always larger for the cross-linked substrate. We attribute this feature the different stiffness: the deformation upon the impact is larger in the case of the EM-PDMS substrate thus resulting in a comparatively reduced increase of temperature.
\begin{figure}[h]
\includegraphics{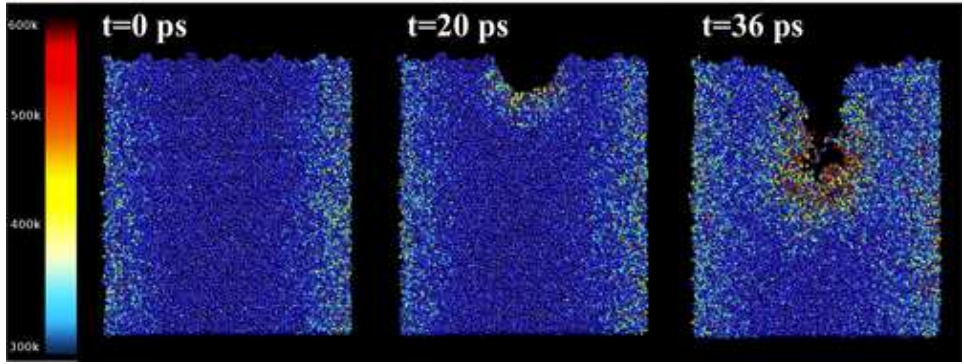}
\caption{Time evolution of the temperature wave on CL-PDMS  during the implantation of a at 2.0 eV/atom.
Each atom is colored according to its local temperature: red color corresponds to a temperature of 600 K while blue color corresponds to 300K. The cluster is not shown for sake of clarity.\label{PD-WAVE}}
\end{figure}
\begin{figure}[h!]
\includegraphics{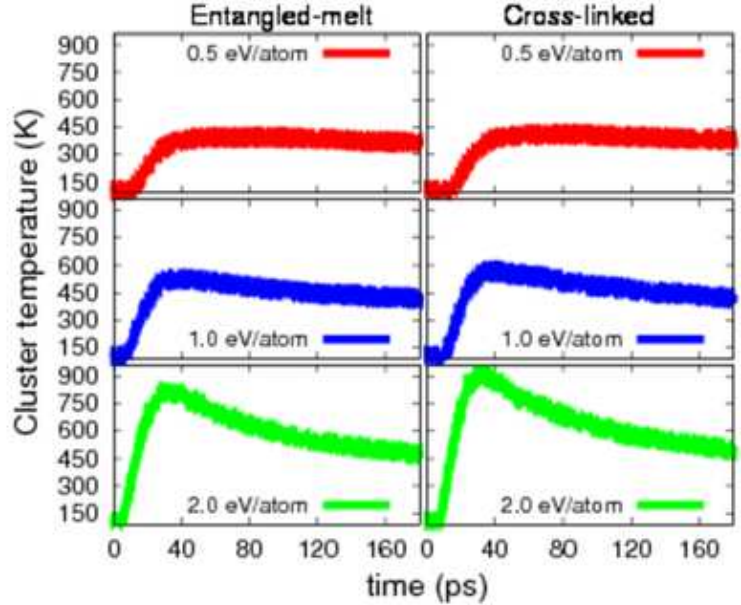}
\caption{Time evolution of the cluster temperature during SCBI simulation for the EM-(left) and CL-(right) PDMS for the implantation energies of 0.5 eV/atom (top), 1.0 eV/atom (middle) and 2.0 eV/atom (bottom).\label{PD-cl}}
 \end{figure}
 
\ref{PD-WAVE} shows the time evolution of the temperature wave on CL-PDMS  for implantation energy of 2.0 eV/atom.
For sake of clarity we represent only a PDMS slice having thickness of 2 nm and height of 30 nm, without the cluster.  Each atom is colored according to its local kinetic temperature: red color corresponds to a temperature of 600 K while blue color corresponds to 300 K.
As due to the cluster impact (for t $\sim$ 20 ps) we observe that the local temperature of the atoms in contact with the implanted Au-nc increases up to $\sim$600 K. This temperature wave propagates inside the polymer  being increasingly  dumped  with the polymer depth.
Therefore, besides the temperature increase observed on the PDMS surface, we observe a dramatic temperature rising on the cluster impact area.  We expect that the local properties of the polymer (such as density, number of covalent bonds, pressure) will be remarkably changed upon  the cluster impact, favoring the penetration of the next clusters to be implanted.
%After $\sim$ $40$ ps, corresponding to the S1 regime previously described, we observe that the temperature wave depth is greater that the actual cluster penetration depth.
 \begin{figure}[h!]
\includegraphics{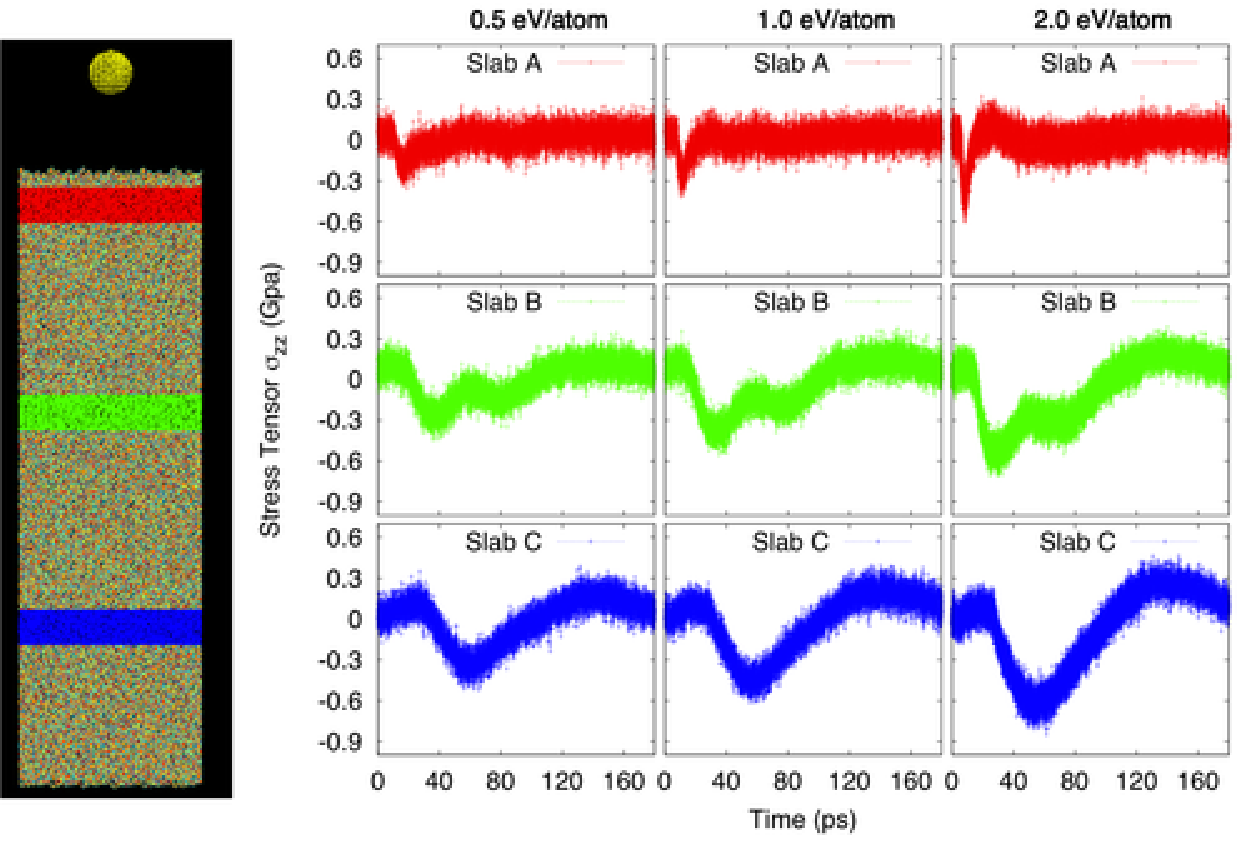}
\caption{ Time evolution of the stress tensor component $z$ ($\sigma_{zz}$) along the implantation direction  during SCBI simulation for the EM-PDMS. Right panel: the first, second and  third  lines  represents the $\sigma_{zz}$ evolutions inside the $A$, $B$ and $C$ slabs respectively. Similar results were found for CL-PDMS\label{PD5}}
\end{figure} 
 
 We also calculate the cluster temperature during implantation  which suddenly increases after the impact as reported in \ref{PD-cl}.
 In \ref{sub2} we summarize the temperature increase $\Delta T_{cl}$ at $t_{end}$  which depends on the implantation energy. The most interesting observation is that the structure of the substrate affects the maximum temperature reached at the cluster.
  \begin{table}[h!]
\begin{tabular}{|c|c|c|c|}
\hline
& 0.5 eV/atom&1.0 eV/atom&2.0 eV/atom\\
\hline
$EM-\Delta T_{cl}$ (K)& 257 & 293  &  369 \\
$CL-\Delta T_{cl}$ (K)& 264 & 307  &  376 \\
\hline
\end{tabular}
\caption{Cluster temperature increase  $\Delta T_{cl}$ at the end of the implantation on EM- and CL-PDMS samples.\label{sub2}}
\end{table}
 % which suddenly increases after the impact as reported in \ref{PD5}.
%$\Delta T_{cl}$ at $t_{end}$  which depends on the implantation energy. For EM-PDMS  $\Delta T_{cl}$ $\sim$50 K for $E_{impl}$=0.5 eV/atom, $\Delta T_{cl}$ $\sim$100 K for $E_{impl}$1.0 eV/atom and  $\Delta T_{cl}$ $\sim$ 150 K for $E_{impl}$=2.0 eV/atom. For CL-PDMS  $\Delta T_{cl}$ $\sim$70 K for $E_{impl}$=0.5 eV/atom, $\Delta T_{cl}$ $\sim$130 K for $E_{impl}$1.0 eV/atom and  $\Delta T_{cl}$ $\sim$ 170 K for $E_{impl}$=2.0 eV/atom.  Therefore we observe that that the structure of the substrate affects the maximum temperature reached at the cluster.
 
The cluster implantation gives rise as well to a pressure wave inside the PDMS.
\ref{PD5} shows the component of the stress tensor $z$ ($\sigma_{zz}$ along the implantation direction of  three slabs ($A$,$B$,$C$ )  during the 180 ps -long the simulation for the EM-PDMS sample. We choose slab $A$ close to the EM-PDMS surface(70-20 nm depth), slab $B$ corresponding the middle of the substrate (370-320 nm depth), and slab $C$ close to the  PDMS bottom (670-620 nm depth). 
 As found for the temperature we observe a pressure wave traveling inside the PDMS substrate upon the cluster implantation. 
From \ref{PD5} we can also identify the same $S_1-S_4$ steps previously detected: in the first part of the simulation ($S_1$) before the cluster impact the pressure oscillates around a constant value. Then, upon the impact ($S_2$) we observe a pressure increase on the direction of the cluster initial velocity. Then, we observe the PDMS response ($S_3$)  marked by a pressure increase on the opposite direction with respect to the cluster initial velocity. Finally we observe the $S_4$ regime when the the pressure stabilizes around the initial value. Similar results were found for CL-PDMS.

 %The cluster penetration depths observed in our molecular dynamics simulation are in the order of tens of nm while the experimental penetration depths are in the order of hundreds of nm \cite{Corbelli2011}. We demonstrate that even a single cluster impact on the PDMS substrate remarkably changes the polymer properties in terms of PDMS local temperature and local pressure. Moreover, we observe the presence  of craters created at the surface  having lateral dimensions comparable to the cluster radius and depths strongly dependent on the implantation energy.  We expect that all these variations on the PDMS morphology  (density decrease, interchain bond breaking, presence of craters) will strongly favor the penetration of the next clusters to be implanted possibly reaching penetration depths in the order of hundreds  of nm at the end of the implantation process.
\subsection{Conclusions}
In the present investigation we study by means of atomistic model potential molecular dynamics the implantation of Au-nanoclusters on a PDMS polymer and the following polymer microstructure evolution. We compare the response of two different PDMS amorphous structures: i) entangled melt and ii) crossed-linked. 
We show that for both entangled-melt and cross-linked  Polydimethylsiloxane the cluster penetration depth inside the polymer linearly depends on the implantation energy having an angular coefficient of $7$ nm/eV in the case of entangled-melt PDMS and of $6$ nm/eV for cross-linked PDMS. In particular we observe that in the case of entangled-melt Polydimethylsiloxane the penetration depth is larger than cross-linked.
We show that even a single cluster impact on the Polydimethylsiloxane substrate remarkably changes the polymer local temperature and pressure. Moreover we observe the presence  of craters created on the polymer surface having lateral dimensions comparable to the cluster radius and depths strongly dependent on the implantation energy.  
Finally, present simulations suggest that the substrate morphology is largely affected by the cluster impact and that most-likely such modifications favor the the penetration of the next  impinging clusters. 
\subsection{Aknowledgments}
This work has been funded by the Regione Autonoma della Sardegna and Regione  Lombardia under Project  "ELDABI''. We acknowledge computational support by CINECA (Casalecchio di Reno, Italy). We are glad to acknowledge Paolo Milani (University of Milan), Luca Ravagnan (WISE S.r.l.), Sandro Ferrari (WISE S.r.l.) and Cristian Ghisleri (WISE S.r.l) for useful discussions.
\bibliography{bibliography}

\begin{thebibliography}{6}%
\makeatletter
\providecommand \@ifxundefined [1]{%
 \@ifx{#1\undefined}
}%
\providecommand \@ifnum [1]{%
 \ifnum #1\expandafter \@firstoftwo
 \else \expandafter \@secondoftwo
 \fi
}%
\providecommand \@ifx [1]{%
 \ifx #1\expandafter \@firstoftwo
 \else \expandafter \@secondoftwo
 \fi
}%
\providecommand \natexlab [1]{#1}%
\providecommand \enquote  [1]{``#1''}%
\providecommand \bibnamefont  [1]{#1}%
\providecommand \bibfnamefont [1]{#1}%
\providecommand \citenamefont [1]{#1}%
\providecommand \href@noop [0]{\@secondoftwo}%
\providecommand \href [0]{\begingroup \@sanitize@url \@href}%
\providecommand \@href[1]{\@@startlink{#1}\@@href}%
\providecommand \@@href[1]{\endgroup#1\@@endlink}%
\providecommand \@sanitize@url [0]{\catcode `\\12\catcode `\$12\catcode
  `\&12\catcode `\#12\catcode `\^12\catcode `\_12\catcode `\%12\relax}%
\providecommand \@@startlink[1]{}%
\providecommand \@@endlink[0]{}%
\providecommand \url  [0]{\begingroup\@sanitize@url \@url }%
\providecommand \@url [1]{\endgroup\@href {#1}{\urlprefix }}%
\providecommand \urlprefix  [0]{URL }%
\providecommand \Eprint [0]{\href }%
\providecommand \doibase [0]{http://dx.doi.org/}%
\providecommand \selectlanguage [0]{\@gobble}%
\providecommand \bibinfo  [0]{\@secondoftwo}%
\providecommand \bibfield  [0]{\@secondoftwo}%
\providecommand \translation [1]{[#1]}%
\providecommand \BibitemOpen [0]{}%
\providecommand \bibitemStop [0]{}%
\providecommand \bibitemNoStop [0]{.\EOS\space}%
\providecommand \EOS [0]{\spacefactor3000\relax}%
\providecommand \BibitemShut  [1]{\csname bibitem#1\endcsname}%
\let\auto@bib@innerbib\@empty
\bibitem[Lee et~al.(2006)Lee, Koo, Park, Moon, Hahn, and Kim]{Lee2006}
Lee,~S.; Koo,~B.; Park,~J.~G.; Moon,~H.; Hahn,~J.; Kim,~J.~M. \emph{Mrs
  Bulletin} \textbf{2006}, \emph{31}, 455--459\relax
%\EndOfBibitem
\bibitem[Rogers et~al.(2001)Rogers, Bao, Baldwin, Dodabalapur, Crone, Raju,
  Kuck, Katz, Amundson, Ewing, and Drzaic]{Rogers2001}
Rogers,~J.~A.; Bao,~Z.; Baldwin,~K.; Dodabalapur,~A.; Crone,~B.; Raju,~V.~R.;
  Kuck,~V.; Katz,~H.; Amundson,~K.; Ewing,~J.; Drzaic,~P. \emph{Proceedings of
  the National Academy of Sciences of the United States of America}
  \textbf{2001}, \emph{98}, 4835--4840\relax
\bibitem[Gelinck et~al.(2004)Gelinck, Huitema, Van~Veenendaal, Cantatore,
  Schrijnemakers, Van~der Putten, Geuns, Beenhakkers, Giesbers, Huisman,
  Meijer, Benito, Touwslager, Marsman, Van~Rens, and De~Leeuw]{Gelinck2004}
Gelinck,~G.~H. et~al.  \emph{Nature Materials} \textbf{2004}, \emph{3},
  106--110\relax
\bibitem[Someya et~al.(2005)Someya, Kato, Sekitani, Iba, Noguchi, Murase,
  Kawaguchi, and Sakurai]{Someya2005}
Someya,~T.; Kato,~Y.; Sekitani,~T.; Iba,~S.; Noguchi,~Y.; Murase,~Y.;
  Kawaguchi,~H.; Sakurai,~T. \emph{Proceedings of the National Academy of
  Sciences of the United States of America} \textbf{2005}, \emph{102},
  12321--12325\relax
\bibitem[Baude et~al.(2003)Baude, Ender, Haase, Kelley, Muyres, and
  Theiss]{Baude2003}
Baude,~P.~F.; Ender,~D.~A.; Haase,~M.~A.; Kelley,~T.~W.; Muyres,~D.~V.;
  Theiss,~S.~D. \emph{Applied Physics Letters} \textbf{2003}, \emph{82},
  3964--3966\relax
\bibitem[Cantatore et~al.(2007)Cantatore, Geuns, Gelinck, van Veenendaal,
  Gruijthuijsen, Schrijnemakers, Drews, and de~Leeuw]{Cantatore2007}
Cantatore,~E.; Geuns,~T. C.~T.; Gelinck,~G.~H.; van Veenendaal,~E.;
  Gruijthuijsen,~A. F.~A.; Schrijnemakers,~L.; Drews,~S.; de~Leeuw,~D.~M.
  \emph{Ieee Journal of Solid-state Circuits} \textbf{2007}, \emph{42},
  IEEE\relax
\bibitem[Ouyang et~al.(2004)Ouyang, Chu, Szmanda, Ma, and Yang]{Ouyang2004}
Ouyang,~J.~Y.; Chu,~C.~W.; Szmanda,~C.~R.; Ma,~L.~P.; Yang,~Y. \emph{Nature
  Materials} \textbf{2004}, \emph{3}, 918--922\relax
\bibitem[Mayer et~al.(2007)Mayer, Scully, Hardin, Rowell, and
  McGehee]{Mayer2007}
Mayer,~A.~C.; Scully,~S.~R.; Hardin,~B.~E.; Rowell,~M.~W.; McGehee,~M.~D.
  \emph{Materials Today} \textbf{2007}, \emph{10}, 28--33\relax
\bibitem[Schubert and Werner(2006)Schubert, and Werner]{Schubert2006}
Schubert,~M.~B.; Werner,~J.~H. \emph{Materials Today} \textbf{2006}, \emph{9},
  42--50\relax
\bibitem[Engel et~al.(2006)Engel, Chen, and Liu]{Engel2006}
Engel,~J.; Chen,~J.; Liu,~C. \emph{Applied Physics Letters} \textbf{2006},
  \emph{89}, 221907\relax
\bibitem[Rosset et~al.(2009)Rosset, Niklaus, Dubois, and Shea]{Rosset2009}
Rosset,~S.; Niklaus,~M.; Dubois,~P.; Shea,~H.~R. \emph{Journal of
  Microelectromechanical Systems} \textbf{2009}, \emph{18}, 1300--1308\relax
\bibitem[Kim et~al.(2008)Kim, Ahn, Choi, Kim, Kim, Song, Huang, Liu, Lu, and
  Rogers]{Kim2008}
Kim,~D.-H.; Ahn,~J.-H.; Choi,~W.~M.; Kim,~H.-S.; Kim,~T.-H.; Song,~J.;
  Huang,~Y.~Y.; Liu,~Z.; Lu,~C.; Rogers,~J.~A. \emph{Science} \textbf{2008},
  \emph{320}, 507--511\relax
\bibitem[Whitesides(2006)]{Whitesides2006}
Whitesides,~G.~M. \emph{Nature} \textbf{2006}, \emph{442}, 368--373\relax
\bibitem[Mandlik et~al.(2008)Mandlik, Gartside, Han, Cheng, Wagner, Silvernail,
  Ma, Hack, and Brown]{Mandlik2008}
Mandlik,~P.; Gartside,~J.; Han,~L.; Cheng,~I.-C.; Wagner,~S.;
  Silvernail,~J.~A.; Ma,~R.-Q.; Hack,~M.; Brown,~J.~J. \emph{Applied Physics
  Letters} \textbf{2008}, \emph{92}, 103309\relax
\bibitem[Dollase et~al.(2002)Dollase, Spiess, Gottlieb, and
  Yerushalmi-Rozen]{Dollase2002}
Dollase,~T.; Spiess,~H.~W.; Gottlieb,~M.; Yerushalmi-Rozen,~R.
  \emph{Europhysics Letters} \textbf{2002}, \emph{60}, 390--396\relax
\bibitem[Graz et~al.(2009)Graz, Cotton, and Lacour]{Graz2009}
Graz,~I.~M.; Cotton,~D. P.~J.; Lacour,~S.~P. \emph{Applied Physics Letters}
  \textbf{2009}, \emph{94}, 071902\relax
\bibitem[Maggioni et~al.(2004)Maggioni, Vomiero, Carturan, Scian, Mattei,
  Bazzan, Fernandez, Mazzoldi, Quaranta, and Della~Mea]{Maggioni2004}
Maggioni,~G.; Vomiero,~A.; Carturan,~S.; Scian,~C.; Mattei,~G.; Bazzan,~M.;
  Fernandez,~G.~D.; Mazzoldi,~P.; Quaranta,~A.; Della~Mea,~G. \emph{Applied
  Physics Letters} \textbf{2004}, \emph{85}, 5712--5714\relax
\bibitem[Ravagnan et~al.(2009)Ravagnan, Divitini, Rebasti, Marelli, Piseri, and
  Milani]{Ravagnan2009}
Ravagnan,~L.; Divitini,~G.; Rebasti,~S.; Marelli,~M.; Piseri,~P.; Milani,~P.
  \emph{Journal of Physics D-applied Physics} \textbf{2009}, \emph{42},
  082002\relax
\bibitem[Corbelli et~al.(2011)Corbelli, Ghisleri, Marelli, Milani, and
  Ravagnan]{Corbelli2011}
Corbelli,~G.; Ghisleri,~C.; Marelli,~M.; Milani,~P.; Ravagnan,~L.
  \emph{Advanced Materials} \textbf{2011}, \emph{23}, 4504\relax
\bibitem[Anders and Urbassek(2008)Anders, and Urbassek]{CLW}
Anders,~C.; Urbassek,~H.~M. \emph{Nuclear Instruments \& Methods In Physics
  Research Section B-beam Interactions With Materials and Atoms} \textbf{2008},
  \emph{266}, 44--48\relax
\bibitem[Carroll et~al.(2000)Carroll, Nellist, Palmer, Hobday, and Smith]{CLW1}
Carroll,~S.~J.; Nellist,~P.~D.; Palmer,~R.~E.; Hobday,~S.; Smith,~R.
  \emph{Physical Review Letters} \textbf{2000}, \emph{84}, 2654--2657\relax
\bibitem[Sanz-Navarro et~al.(2002)Sanz-Navarro, Smith, Kenny, Pratontep, and
  Palmer]{CLW2}
Sanz-Navarro,~C.~F.; Smith,~R.; Kenny,~D.~J.; Pratontep,~S.; Palmer,~R.~E.
  \emph{Physical Review B} \textbf{2002}, \emph{65}, 165420\relax
\bibitem[Marelli et~al.(2011)Marelli, Divitini, Collini, Ravagnan, Corbelli,
  Ghisleri, Gianfelice, Lenardi, Milani, and Lorenzelli]{Marelli2011}
Marelli,~M.; Divitini,~G.; Collini,~C.; Ravagnan,~L.; Corbelli,~G.;
  Ghisleri,~C.; Gianfelice,~A.; Lenardi,~C.; Milani,~P.; Lorenzelli,~L.
  \emph{Journal of Micromechanics and Microengineering} \textbf{2011},
  \emph{21}, 045013\relax
\bibitem[Sun(1998)]{Sun1998}
Sun,~H. \emph{Journal of Physical Chemistry B} \textbf{1998}, \emph{102},
  7338--7364\relax
\bibitem[Heinz et~al.(2008)Heinz, Vaia, Farmer, and Naik]{Heinz2008}
Heinz,~H.; Vaia,~R.~A.; Farmer,~B.~L.; Naik,~R.~R. \emph{Journal of Physical
  Chemistry C} \textbf{2008}, \emph{112}, 17281--17290\relax
\bibitem[Jonsdottir and Rasmussen(2000)Jonsdottir, and Rasmussen]{CFF}
Jonsdottir,~S.~O.; Rasmussen,~K. \emph{New Journal of Chemistry} \textbf{2000},
  \emph{24}, 243--247\relax
\bibitem[Sun and Rigby(1997)Sun, and Rigby]{Sun1997}
Sun,~H.; Rigby,~D. \emph{Spectrochimica Acta Part A-molecular and Biomolecular
  Spectroscopy} \textbf{1997}, \emph{53}, 1301--1323\relax
\bibitem[SUN(1995)]{Sun1995}
SUN,~H. \emph{Macromolecules} \textbf{1995}, \emph{28}, 701--712\relax
\bibitem[Makrodimitri et~al.(2007)Makrodimitri, Dohrn, and
  Economou]{PDMS-AMORPH3}
Makrodimitri,~Z.~A.; Dohrn,~R.; Economou,~I.~G. \emph{Macromolecules}
  \textbf{2007}, \emph{40}, 1720\relax
\bibitem[Sides et~al.(2002)Sides, Curro, Grest, Stevens, Soddemann,
  Habenschuss, and London]{PDMS-AMORPH4}
Sides,~S.~W.; Curro,~J.; Grest,~G.~S.; Stevens,~M.~J.; Soddemann,~T.;
  Habenschuss,~A.; London,~J.~D. \emph{Macromolecules} \textbf{2002},
  \emph{35}, 6455\relax
\bibitem[Theodorou and Suter(1985)Theodorou, and Suter]{AMORPH}
Theodorou,~D.~N.; Suter,~U.~W. \emph{Macromolecules} \textbf{1985}, \emph{18},
  1485\relax
\bibitem[Theodorou and Suter(1986)Theodorou, and Suter]{AMORPH1}
Theodorou,~D.~N.; Suter,~U.~W. \emph{Macromolecules} \textbf{1986}, \emph{19},
  139\relax
\bibitem[MAT()]{MATERIALSTUDIO}
http://accelrys.com/products/materials-studio/\relax
\bibitem[Fakhreddine and Zoller(1990)Fakhreddine, and Zoller]{PDMS-AMORPH1}
Fakhreddine,~Y.~A.; Zoller,~P. \emph{J. Appl. Polym. Sci.} \textbf{1990},
  \emph{41}, 1087\relax
\bibitem[Shih and J.Flory(1972)Shih, and J.Flory]{PDMS-AMORPH2}
Shih,~H.; J.Flory,~P. \emph{Macromolecules} \textbf{1972}, \emph{5}, 758\relax
\bibitem[Chenoweth et~al.(2005)Chenoweth, Cheung, van Duin, Goddard, and
  Kober]{Chenoweth2005}
Chenoweth,~K.; Cheung,~S.; van Duin,~A. C.~T.; Goddard,~W.~A.; Kober,~E.~M.
  \emph{Journal of the American Chemical Society} \textbf{2005}, \emph{127},
  7192--7202\relax
\bibitem[Shemella et~al.(2011)Shemella, Laino, Fritz, and
  Curioni]{Shemella2011}
Shemella,~P.~T.; Laino,~T.; Fritz,~O.; Curioni,~A. \emph{Journal of Physical
  Chemistry B} \textbf{2011}, \emph{115}, 2831--2835\relax
\bibitem[Lacevic et~al.(2008)Lacevic, Gee, Saab, and Maxwell]{Lacevic2008}
Lacevic,~N.; Gee,~R.~H.; Saab,~A.; Maxwell,~R. \emph{Journal of Chemical
  Physics} \textbf{2008}, \emph{129}, 124903\relax
\bibitem[Shernella et~al.(2011)Shernella, Laino, Fritz, and
  Curioni]{Shernella2011}
Shernella,~P.~T.; Laino,~T.; Fritz,~O.; Curioni,~A. \emph{Journal of Physical
  Chemistry B} \textbf{2011}, \emph{115}, 13508--13512\relax
\bibitem[Heine et~al.(2004)Heine, Grest, Lorenz, Tsige, and Stevens]{Heine2004}
Heine,~D.~R.; Grest,~G.~S.; Lorenz,~C.~D.; Tsige,~M.; Stevens,~M.~J.
  \emph{Macromolecules} \textbf{2004}, \emph{37}, 3857--3864\relax
\bibitem[MIL()]{MILANI}
Private communications from Paolo Milani\relax
\bibitem[PLIMPTON(1995)]{PLIMPTON1995}
PLIMPTON,~S. \emph{Journal of Computational Physics} \textbf{1995}, \emph{117},
  1--19\relax
\bibitem[Melis et~al.(2013)Melis, L.Colombo, Ravagnan, and Milani]{milani1}
Melis,~C.; L.Colombo,; Ravagnan,~L.; Milani,~P. \textbf{2013}, \relax
% Create the reference section using BibTeX:
%\bibliography{your-bib-file} % render your BibTeX file on your system first, then copy and paste the entire contents of the BBL output file into this file.
\end{thebibliography}
\newpage
%\begin{figure}[h!]
%\includegraphics[scale=0.2]{10.eps}
%\caption{Table of Contents Graphic, "Neutral-cluster implantation in polymers by computer experiments", R. Cardia, C. Melis and L. Colombo}
%\end{figure} 
% If in two-column mode, this environment will change to single-column format so that long equations can be displayed. 
% Use only when necessary.
%\begin{widetext}
%$$\mbox{put long equation here}$$
%\end{widetext}

% Figures should be put into the text as floats. 
% Use the graphics or graphicx packages (distributed with LaTeX2e).
% See the LaTeX Graphics Companion by Michel Goosens, Sebastian Rahtz, and Frank Mittelbach for examples. 
%
% Here is an example of the general form of a figure:
% Fill in the caption in the braces of the \caption{} command. 
% Put the label that you will use with \ref{} command in the braces of the \label{} command.
%
% \begin{figure}
% \includegraphics{}%
% \caption{\label{}}%
% \end{figure}

% Tables may be be put in the text as floats.
% Here is an example of the general form of a table:
% Fill in the caption in the braces of the \caption{} command. Put the label
% that you will use with \ref{} command in the braces of the \label{} command.
% Insert the column specifiers (l, r, c, d, etc.) in the empty braces of the
% \begin{tabular}{} command.
%
% \begin{table}
% \caption{\label{} }
% \begin{tabular}{}
% \end{tabular}
% \end{table}

% If you have acknowledgments, this puts in the proper section head.
%\begin{acknowledgments}
% Put your acknowledgments here.
%\end{acknowledgments}

\end{document}